\documentclass[preprint,number,sort&compress]{elsarticle}
\usepackage[utf8]{inputenc}
\usepackage{amsmath}
\usepackage{amsfonts}
\usepackage{amssymb}
\usepackage{lineno}
\usepackage{hyperref}
\usepackage{graphicx}

\biboptions{sort&compress}

\begin{document}
	
	\begin{frontmatter}
		
		\title{Simulation studies for dielectric wakefield programme at CLARA facility}

		\author[CI,UoM]{T. H. Pacey\corref{correspond}}
		\cortext[correspond]{Corresponding author}
		\ead{thomas.pacey@cockcroft.ac.uk}
	
		\author[CI,ASTEC]{Y. Saveliev}
		\author[CI,UoM]{G. Xia}
		\author[TechX]{J. Smith}

		\address[CI]{The Cockcroft Institute, Daresbury Laboratory,
		Warrington, UK}
		\address[UoM]{School of Physics and Astronomy, University of Manchester, Manchester, UK}
		\address[ASTEC]{Accelerator Science and Technology Centre, STFC, Daresbury Laboratory, Warrington, UK}
		\address[TechX]{Tech-X UK Ltd, The Innovation Centre, Sci-Tech Daresbury, Warrington, United Kingdom}
		
		\begin{abstract}
			Short, high charge electron bunches can drive high magnitude electric fields in dielectric lined structures. The interaction of the electron bunch with this field has several applications including high gradient dielectric wakefield acceleration (DWA) and passive beam manipulation. The simulations presented provide a prelude to the commencement of an experimental DWA programme at the CLARA accelerator at Daresbury Laboratory. The key goals of this program are: tunable generation of THz radiation, understanding of the impact of transverse wakes, and design of a dechirper for the CLARA FEL. Computations of longitudinal and transverse phase space evolution were made with Impact-T and VSim to support both of these goals.
		\end{abstract}
		
		
	\end{frontmatter}


\section{Introduction}
Dielectric lined waveguides (DLW) are able to support fields in excess of GVm$^{-1}$ \cite{thompson2008breakdown}, that can be excited by short, high charge, relativistic electron bunches. These fields can either accelerate the tail of the driving bunch \cite{Andonian2012} or a smaller trailing witness bunch \cite{o2016observation} in the dielectric wakefield acceleration (DWA) scheme. The wakefield can be coupled out of the structure to generate short, high power pulses of THz radiation. This has been experimentally demonstrated in both dielectric lined and \cite{antipov2016efficient,lekomtsev2017sub} corrugated metallic structures \cite{bane2017measurements,smirnov2015observation}. The wakefield can also be used to passively manipulate the electron bunch, by removing correlated energy spread (dechirping) to improve free electron laser (FEL) performance \cite{craievich2010passive,antipov2014experimental,emma2014experimental,guetg2016commissioning}. If the bunch is offset from the structure axis, the transverse wake causes a longitudinally correlated beam deflection, which can allow for fresh slice two-colour FEL operation \cite{lutman2016fresh,bettoni2016two,zemella2017measurements}, and passive transverse streaking \cite{bettoni2016temporal}. The intrinsic synchronisation of the bunch with the wakefield is the key advantage of these manipulation schemes.

A new experimental facility CLARA is currently under construction at Daresbury Laboratory \cite{clarke2014clara}. This facility aims for developing novel FEL concepts and will also host a diverse exploitation programme providing electron beams with up to 250~MeV beam energy, bunch charges of 250~pC and sub-ps bunch lengths. These beam parameters are sufficient for research of novel acceleration concepts including, for example, plasma wakefield acceleration \cite{xia2016plasma}. CLARA Front End (-FE) has been recently constructed that can deliver the beams with 40-50~MeV beam energies. In combination with the already existing VELA beamline \cite{mcintosh2017vela} and a dedicated experimental test area, this provides a platform for initiation of DWA programme at CLARA. This paper outlines the rationale behind the choice of the first DLW to be investigated and summarises corresponding simulation results. Feasibility of a compact DLW based dechirper for CLARA FEL is also discussed.

\section{DLW Design}
\begin{figure}[h]
	\centering
	\includegraphics[width=0.49\linewidth]{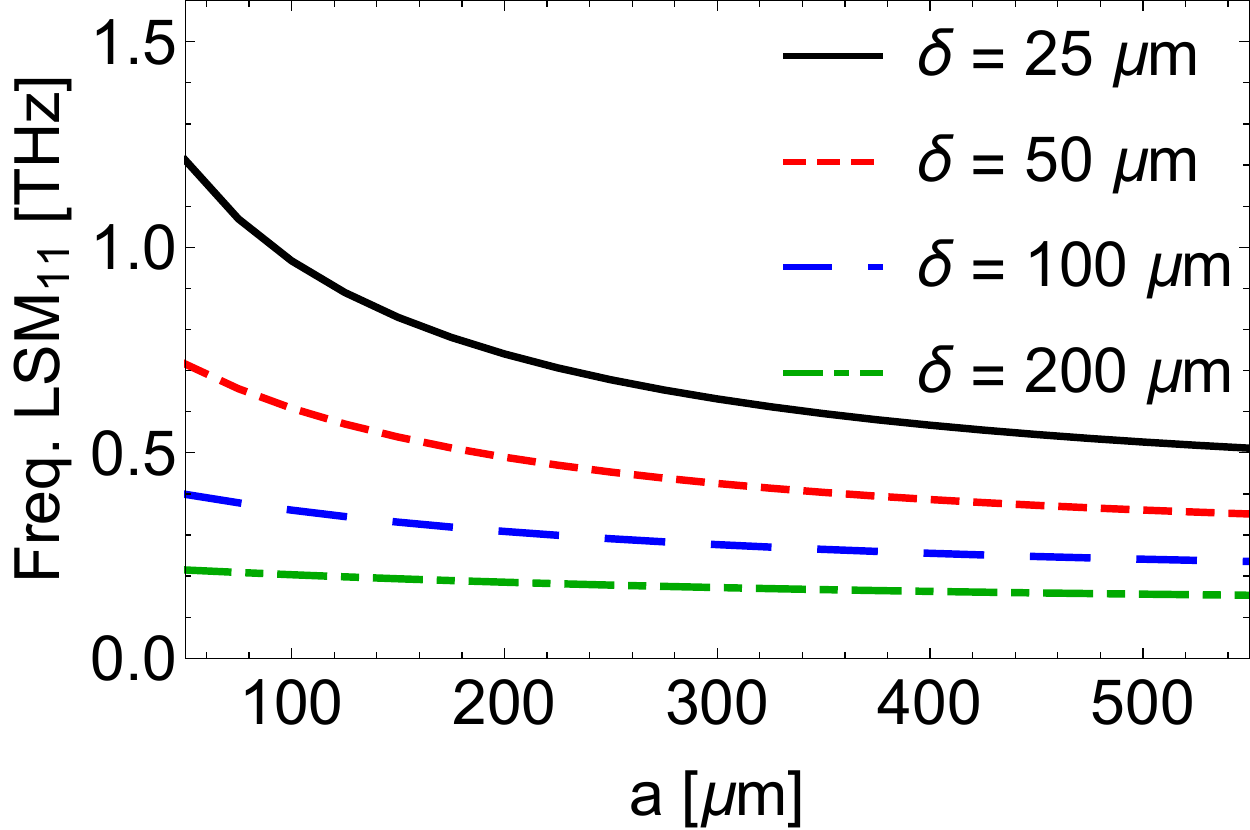}
	\includegraphics[width=0.49\linewidth]{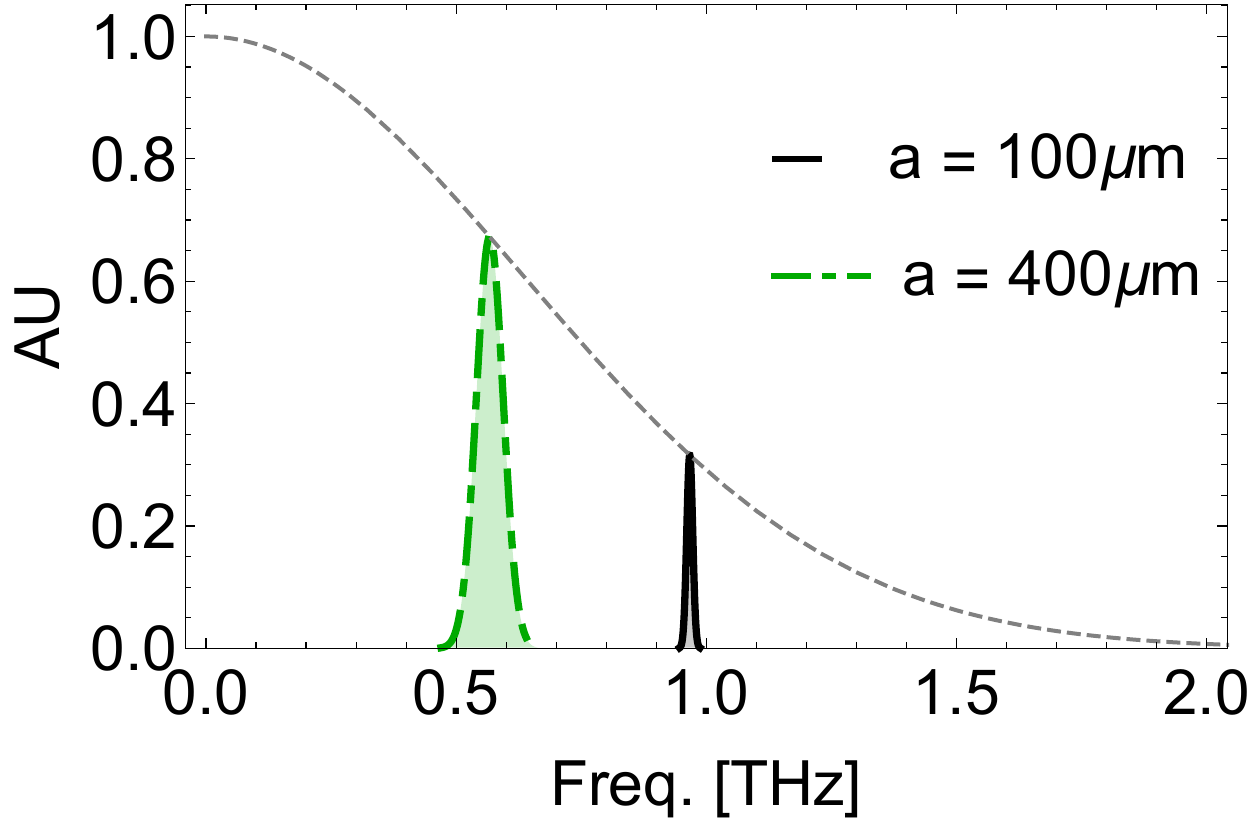}
	\caption{(Colour) Left: Calculated frequency of fundamental wakefield mode as function of both half-aperture and quartz layer thickness. Right: Calculated wakefield spectrum for DLW using parameters in Table \ref{tab:waveguideparams}, normalised to form factor of a Gaussian electron bunch with RMS length $\sigma_t=250$~fs. }
	\label{fig:thinditunability}
\end{figure}

The first phase of the DWA programme aims to investigate tunable THz generation, transverse wakefield effects and to conduct initial dechirping studies for a future CLARA FEL dechirper. These experimental results will also provide a solid base for benchmarking of computer models. A natural choice for that initial programme is a planar structure with variable gap. For given structure dimensions it is possible to calculate the frequencies of longitudinal section magnetic (LSM) and longitudinal section electric (LSE) modes that have phase velocity $v_p=c$ \cite{tremaine1997electromagnetic,mihalcea2012three}. These are the wakefield modes and are analogous to the transverse electric and magnetic modes seen in all metal waveguides. For a waveguide with $a\ll w$ coupling to LSE modes is significantly reduced relative to LSM modes.
The lowest order LSM$_{11}$ mode has a longitudinal electric field component, making this the fundamental mode for acceleration of a witness bunch in a DWA. It should be noted that this mode, shifted by $\pi/2$ in phase, has an electric quadrupole component; this is a result of the boundary conditions and occurs without a drive beam offset \cite{mihalcea2012three}. Therefore transverse wakefields are always excited in a rectangular DLW.
The DLW discussed here is a typical rectangular structure as described in \cite{Andonian2012}, with a tunable half-aperture $a$. The frequency of the fundamental LSM mode is shown in Figure \ref{fig:thinditunability}. This demonstrates that the thinner dielectric layers offer increased frequency tunability over the given half-aperture range.  Specific waveguide parameters are given in Table~\ref{tab:waveguideparams}. Quartz was selected for the dielectric layers due to its availability and ease of manufacture to the required dimensions. 

\begin{table}[h]
	\centering
	\caption{DLW notation and parameters chosen for experiments at CLARA}
	\begin{tabular}{lcc}
		\hline
		Parameter & Symbol & Value \\
		\hline
		Width & $w$ & 2~mm \\
		Length & $L$ & 4~cm \\
		Dielectric thickness & $ \delta$ & 25 $\mu$m \\
		Relative permittivity & $\epsilon_r $ & 3.8 \\
		Half-aperture & $a$ & $(50-500)\mu$m \\
		
		\hline
	\end{tabular}
	
	\label{tab:waveguideparams}
\end{table}

 Application of generalised DLW theory \cite{xiao2001field} allows for calculation of waveguide dispersion relation, and mode group velocity, $v_g$. The pulse length, $\Delta t$, can be approximated using $\Delta t \approx L(c-v_g)/cv_g$, where $c$ is the speed of light \cite{o2016observation}. The pulse bandwidth is then inversely proportional to the pulse length. This allows for calculation of the wakefield spectrum, shown in Figure \ref{fig:thinditunability}. This demonstrates the DLW can produce a single-moded THz frequency output with high tunability. It should be noted that increases in aperture also increase the mode bandwidth, and will have reduced power output. The bandwidth increase is caused by an increase in group velocity; for $a = 100$~$\mu$m $v_g = 0.58c$, and $a = 400$~$\mu$m $v_g = 0.86c$. The thin dielectric layer of $\delta = 25$~$\mu$m was chosen to achieve the experimental aim of short pulse, tunable THz output. Additionally other simulation studies have shown that thin dielectric layers can be used when coupling to short electron pulses, for example those from laser wakefield accelerated bunches \cite{nie2015wakefields,nie2016potential}.

\section{Simulation Results}

Two codes were used for simulation studies: the space charge tracking code Impact-T \cite{qiang2006three} and the particle-in-cell (PIC) code VSim \cite{vorpal}. Impact-T uses an electrostatic PIC model for bunch space charge and analytical mode decomposition method to calculate the DLW wakefield \cite{mihalcea2012three}. This approach offers a significant reduction in simulation time compared to fully electromagnetic PIC methods and is especially useful for structures that are transversely large relative to the driving bunch, for example a dechirper with large aperture or a wide waveguide with a flat beam for DWA. Whilst more computationally expensive, VSim is a more flexible code, allowing for inclusion of additional physics and demonstrated over a wider range of beam parameters. Previously the two codes were compared longitudinally in \cite{mihalcea2012three} and very good agreement was found. Understanding and benchmarking of transverse dynamics is also important for developing DLW beam manipulation applications, for both dechirpers \cite{zhang2015electron} and passive streakers \cite{craievich2017effects}.

Simulations with both computer codes were performed using simulated electron bunch parameters summarised in Table \ref{tab:beamparams}. The bunches were defined using a 6D Gaussian approximation. For clarity in the effects of the wakefields, the bunches were transversely cold and with negligible uncorrelated energy spread. The bunch longitudinal compression to sub-ps levels takes place in the dog-leg section between CLARA-FE and VELA hence the energy chirp introduced in the accelerating linac must be positive, i.e. electrons at the head of the bunch are of higher energy. As the dog-leg section is $\sim 20$~m away from the experimental station, the longitudinal space charge effects increase the correlated energy spread even further resulting in $\Delta E /E \sim 0.5$~\%.  

\begin{table}[h]
	\centering
	\caption{Beam parameters used for simulations of DWA experiment.}
	\begin{tabular}{lcc}
		\hline
		Parameter & Symbol & Value \\
		
		\hline
		Energy & $E$ & 45~MeV \\
		Energy Spread & $\Delta E/E$ & 0.5\% \\
		Charge & $Q$ & 250~pC \\
		RMS length & $\sigma_t$ & 250~fs \\
		RMS width & $\sigma_x$ & 100~$\mu$m \\
		RMS height & $\sigma_y$ & 100~$\mu$m \\
		
		\hline
	\end{tabular}
	
	\label{tab:beamparams}
\end{table}

\subsection{Longitudinal Effects}

\begin{figure*}[h]
	\centering
	\includegraphics[width=0.32\linewidth]{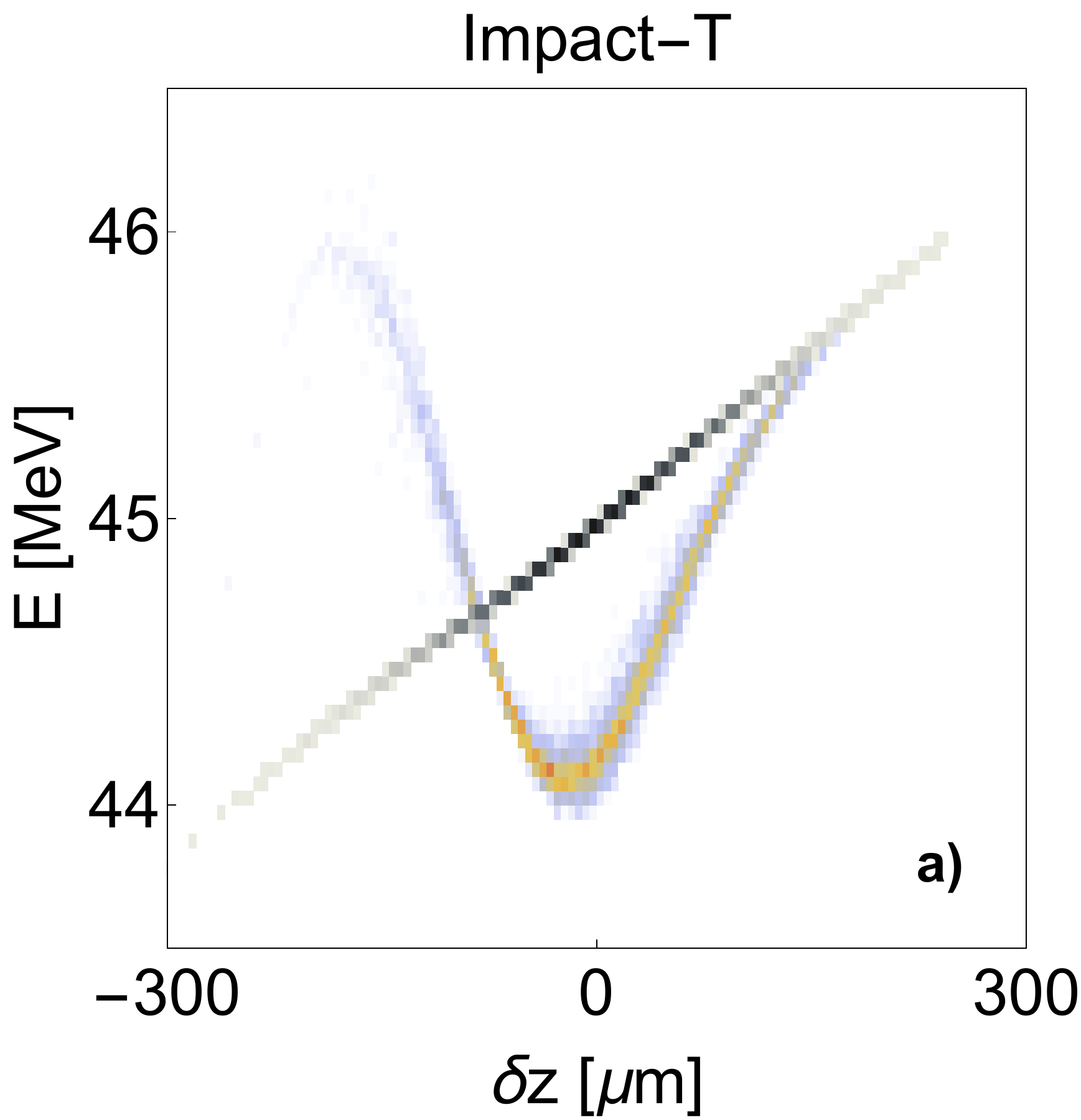}
	\includegraphics[width=0.32\linewidth]{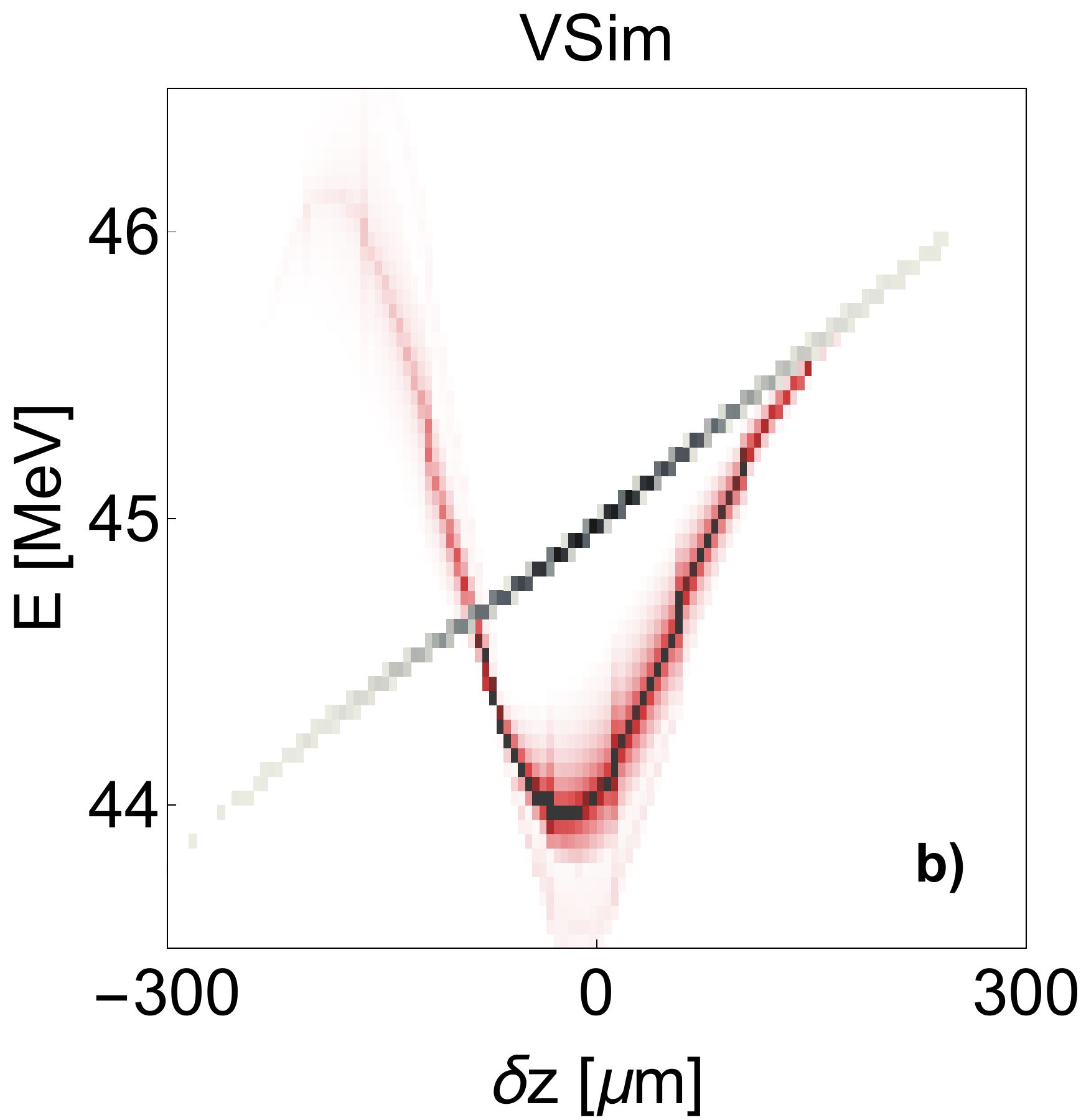}
	\includegraphics[width=0.32\linewidth]{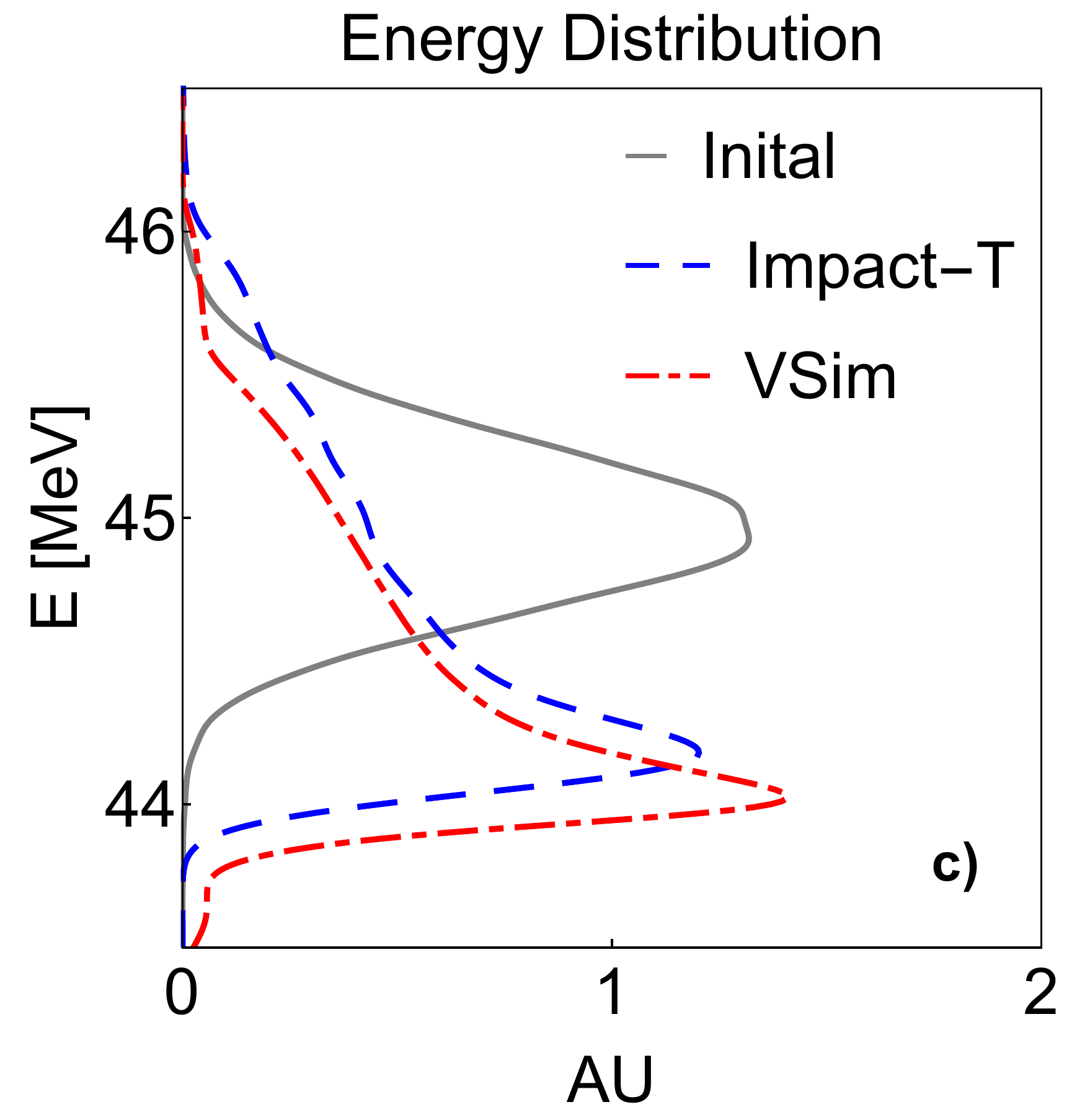}
	\includegraphics[width=0.32\linewidth]{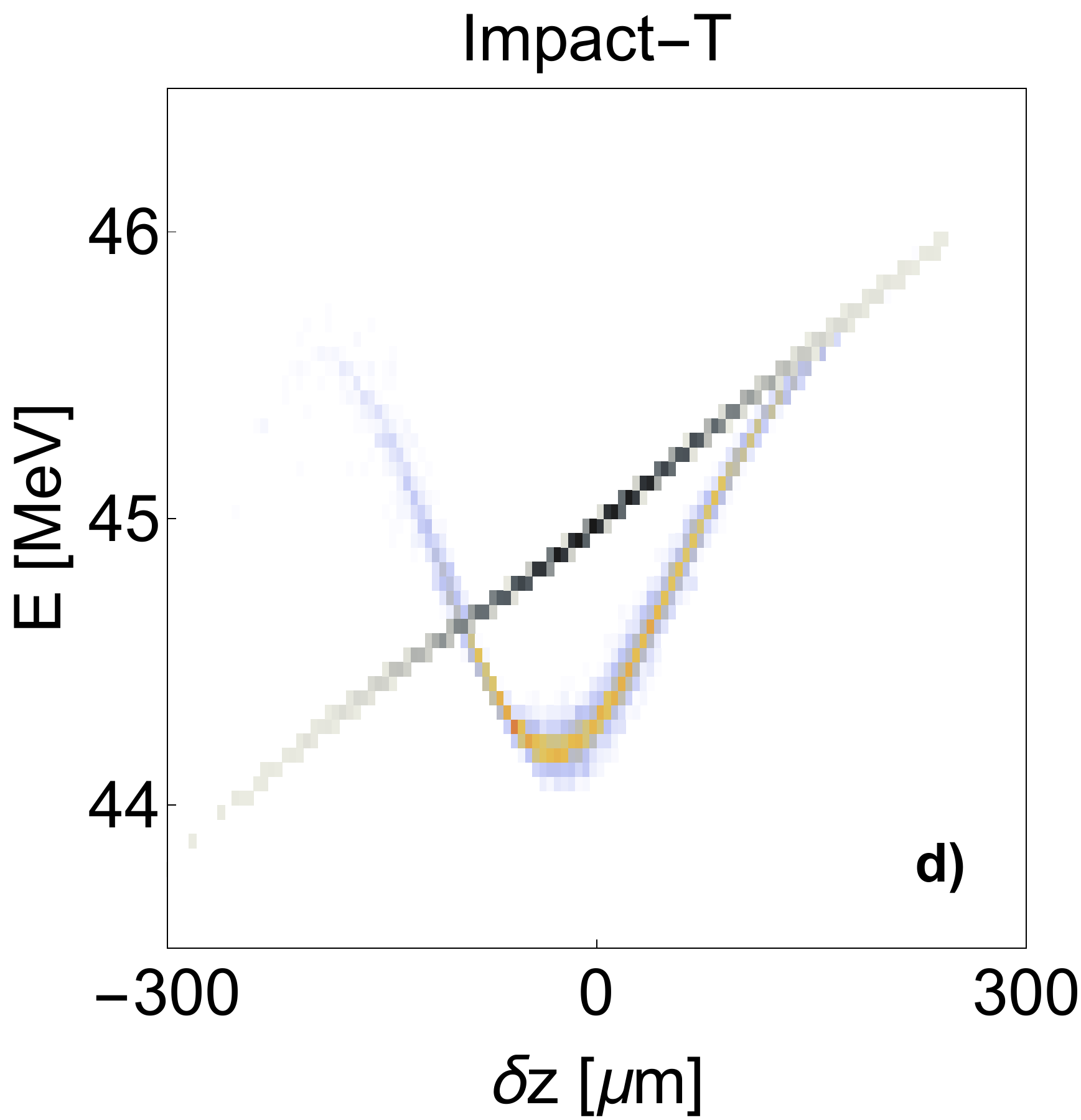}
	\includegraphics[width=0.32\linewidth]{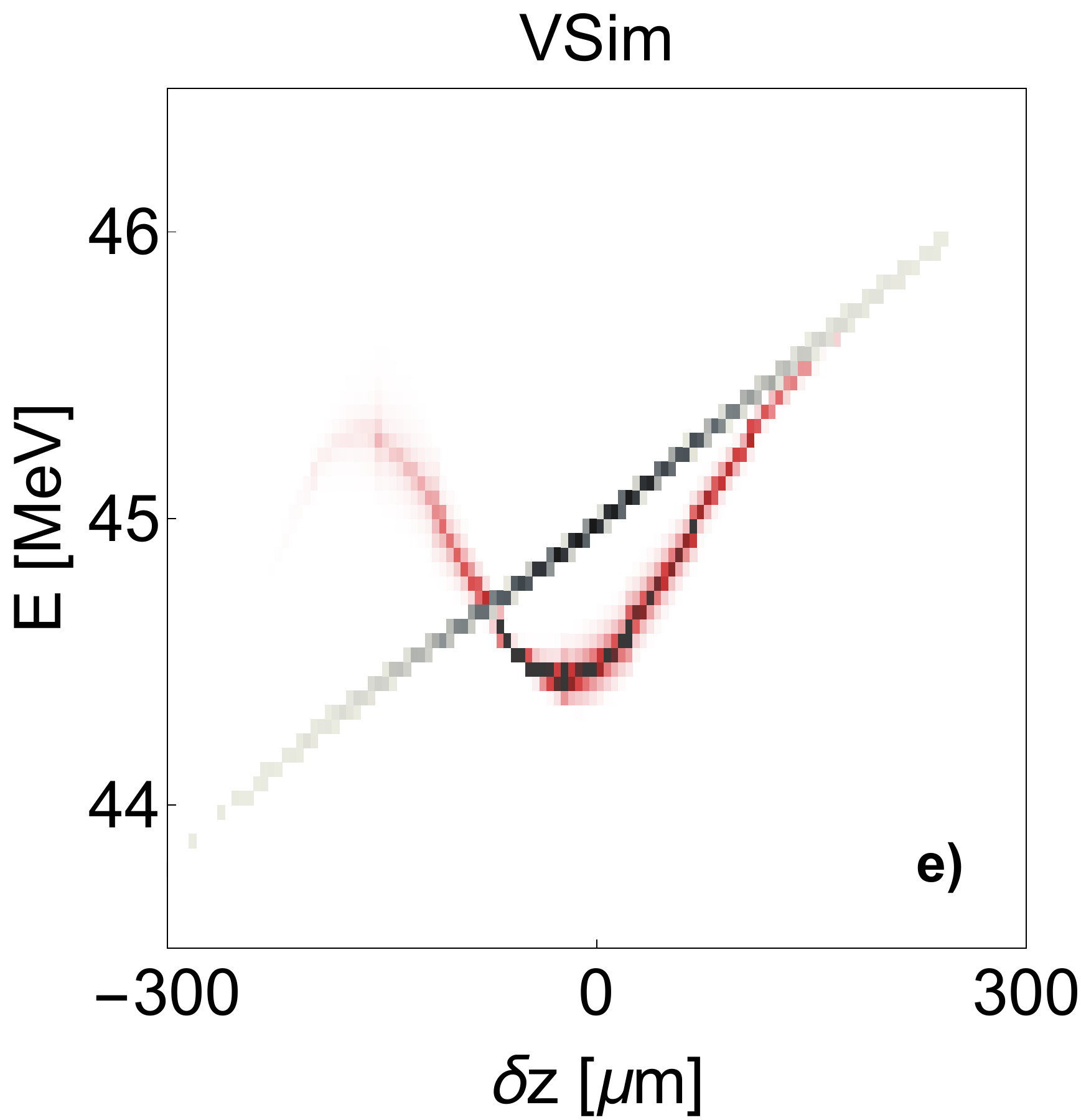}
	\includegraphics[width=0.32\linewidth]{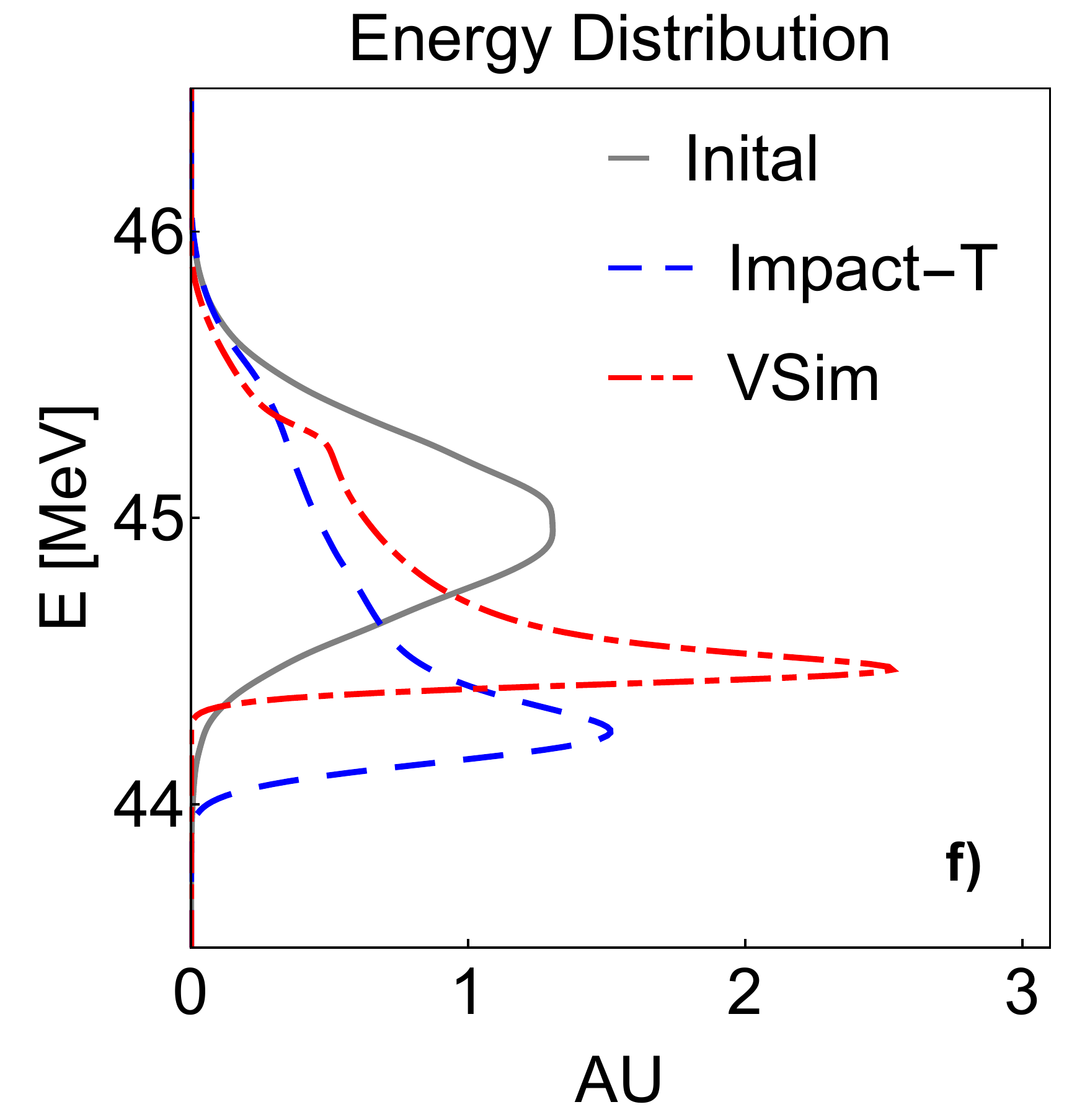}
	\caption{(Colour) Comparison of  simulation results from Impact-T and VSim, with a)-c) $a = 200$~$\mu$m, and  d)-f) $a = 250$~$\mu$m. Normalised density histograms are used for a), b), d), e), with the initial longitudinal phase space with positive chirp is shown in grey, $\delta_z$ is particle coordinate normalised to bunch centroid. }
	\label{fig:LongSims}
\end{figure*}

The thin dielectric layer gives a relatively high frequency wakefield, leading to head-tail acceleration of the electron beam, as seen in the longitudinal phase spaces (LPS) in Figure \ref{fig:LongSims}. There was good qualitative agreement between Impact-T and VSim. In all cases simulated, VSim predicted less energy loss compared to Impact-T, as seen in Figure \ref{fig:LongSims}f). The result for $a = 200 \mu$m (Figure \ref{fig:LongSims}c)) is an exception to this, but the results of the two codes are comparable.  Due to the positive chirp it is not straightforward to differentiate the accelerated tail from the head of the bunch from energy spectra, as in \cite{Andonian2012}. However, the energy distribution of the bunch was altered significantly and uncorrelated (slice) energy spread increased only slightly. Use of a tunable DLW allows for this to be done in a controlled manner, reducing the energy spread of the core of the bunch. For example, in Figure \ref{fig:LongSims}f) the FWHM of the energy spread was reduced from 0.7~MeV to 0.4~MeV in Impact-T and 0.2~MeV in VSim. The variation between the two codes is likely due differences in the final energy spectra; this is a subject of further investigation. Dechirpers typically work for a beam with negative chirp, that is compressed with chicanes, as seen in Section \ref{sec:CLARADechirper}. The tunable DLW with thin dielectric layer can however act as a a `positive dechirper' and this will be a subject of experimental studies.

\subsection{Transverse Effects}

Compared to longitudinal beam dynamics in DLW structures, transverse wakefield effects were given less attention in published literatures. However, these effects are of great importance in any DLW application where preservation of the beam quality is essential, e.g. in single-pass FELs. Transverse effects can be significantly reduced in planar DLWs if flat beam conditions are satisfied, i.e.  $\sigma_x >>\sigma_y$ and $\sigma_x \sim w$ \cite{Andonian2012,tremaine1997electromagnetic,mihalcea2012three}, but such beams may not be suitable for applications like energy dechirpers. It is therefore essential to benchmark computer models against experimental measurements of transverse phase-space for development of practical DLW based solutions. 

\begin{figure}[h]
	\centering
	\includegraphics[width=0.48\linewidth]{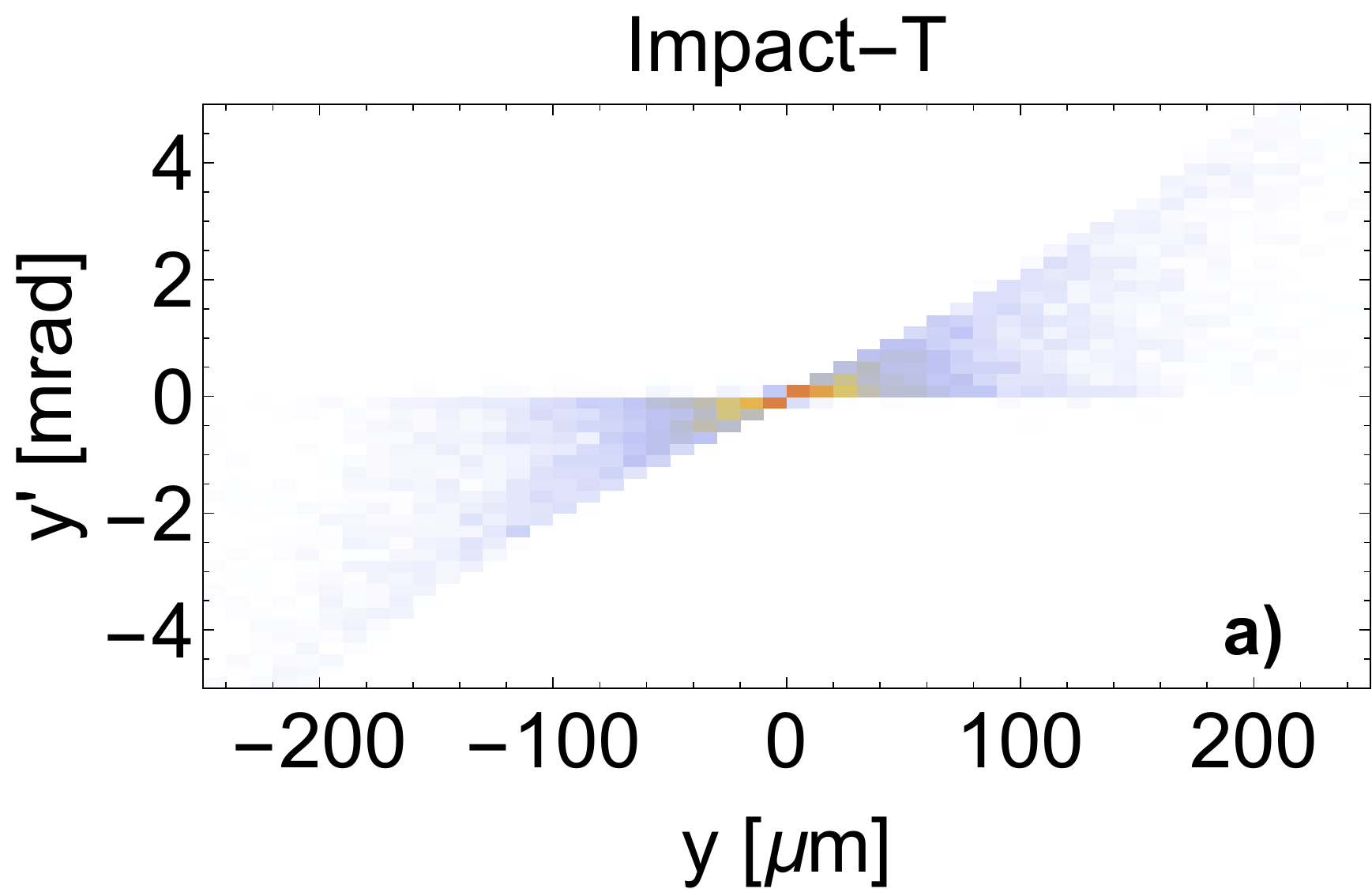}
	\includegraphics[width=0.48\linewidth]{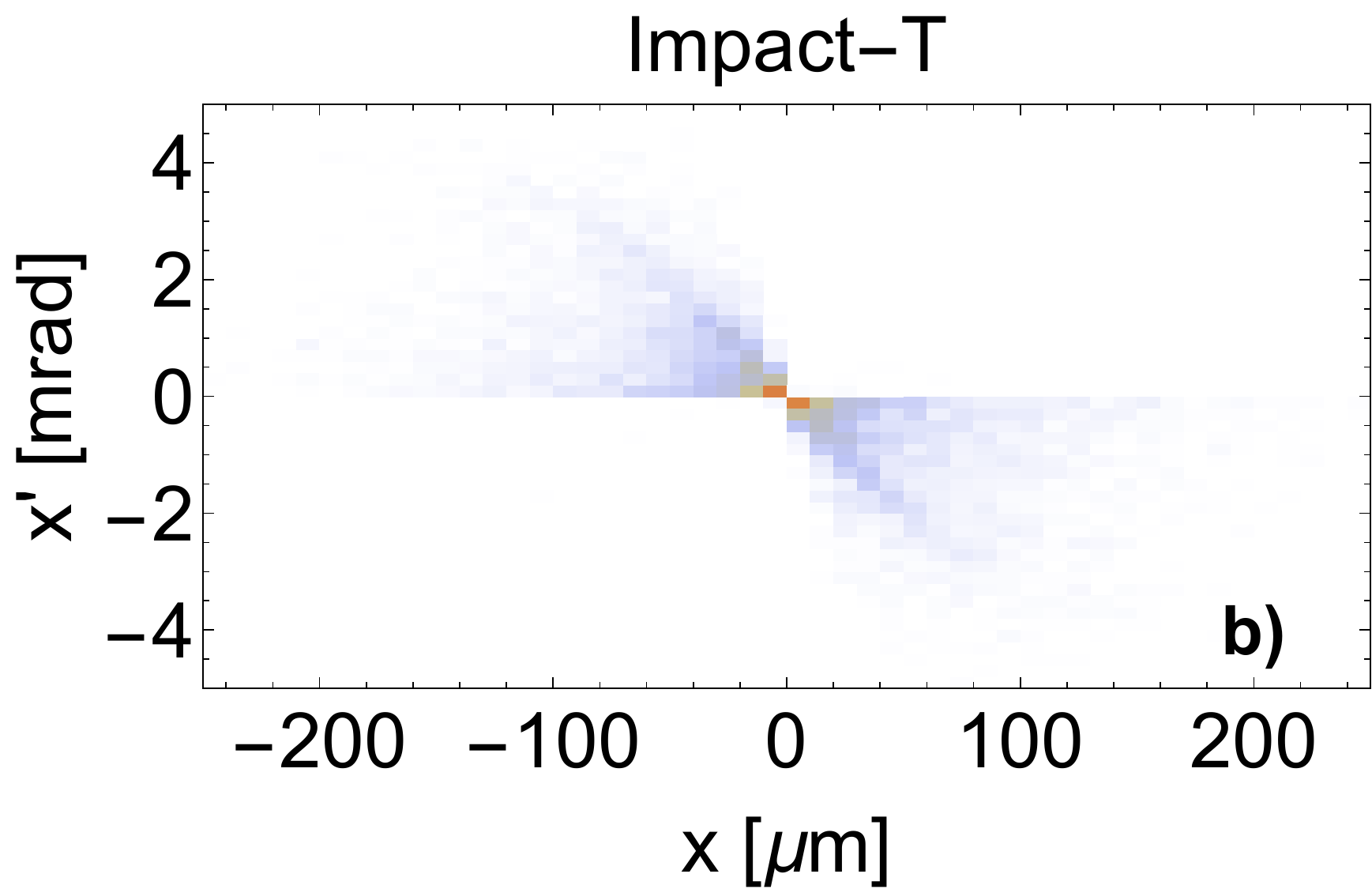}
	\includegraphics[width=0.48\linewidth]{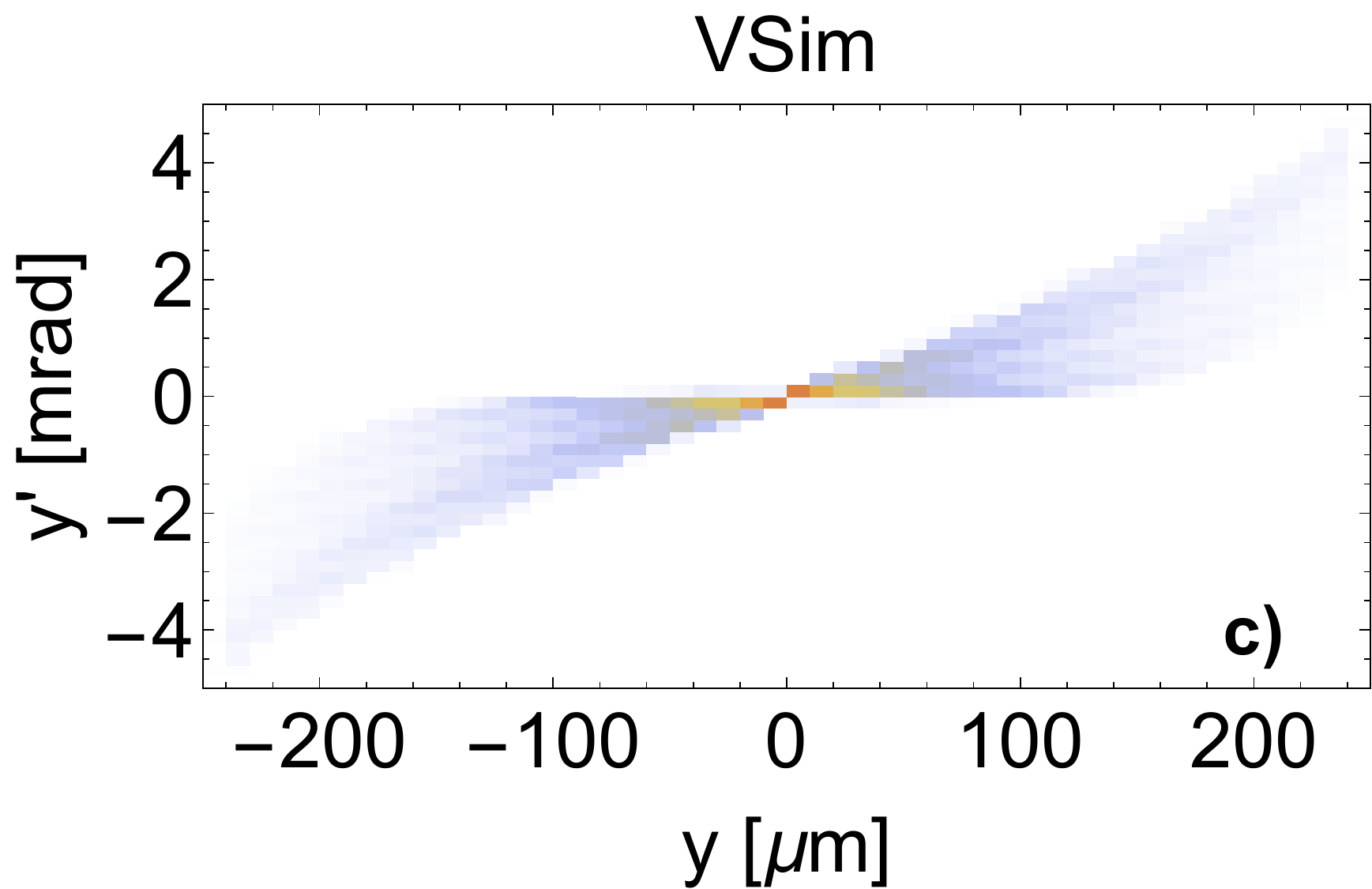}
	\includegraphics[width=0.48\linewidth]{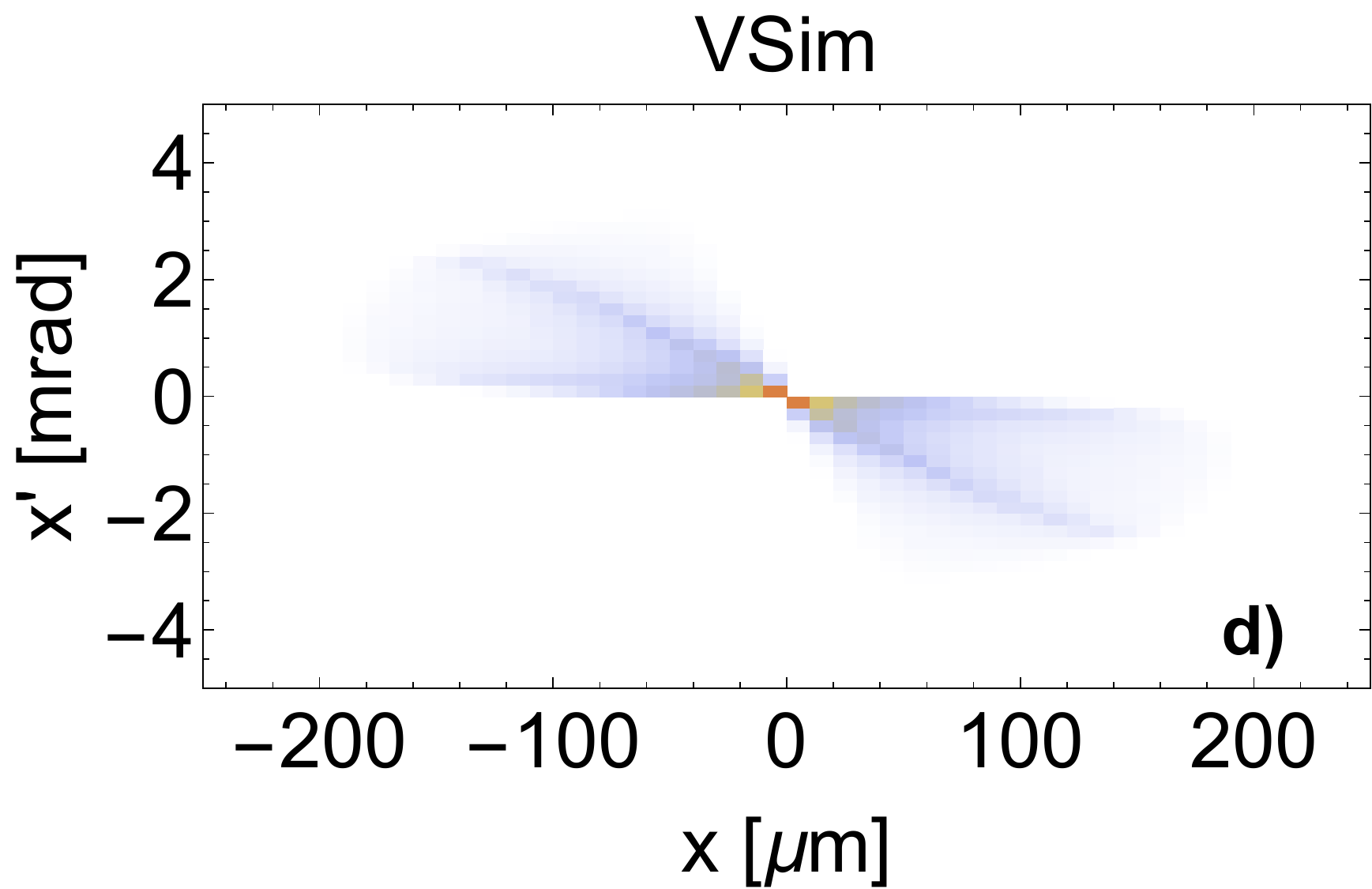}
	\caption{(Colour) Normalised density histograms for transverse phase spaces for $a = 250$~$\mu$m at exit of DLW}
	\label{fig:a250TrsPS}
\end{figure}

For the planar structure and beam parameters given in Tables \ref{tab:waveguideparams} and \ref{tab:beamparams} respectively, transverse phase-spaces were simulated by using both Impact-T and VSim and results for $a = 250$~$\mu$m are presented in Figure \ref{fig:a250TrsPS}.
The transverse wakes observed here are a result of the quadrupole component of the LSM wakefield modes.
 A larger rotation in the $x$ phase space was observed in Impact-T compared to VSim. It should be noted the lower number of fixed weight macro particles used in Impact-T leads to a poorer sampling of the phase space, compared to the large number of variable weight macro particles in VSim. Both codes agree qualitatively in representing transverse phase-spaces. Note a strong beam focusing in horizontal plane and equally strong defocussing in a vertical plane. 
 The Impact-T code gives approximately 50\% higher values for transverse projected emittance in both planes compared to those given by VSim. The values of the projected emittances are not instructive because the slice emittances are not affected by the transverse wakefields \cite{pacey2017phase}; this discrepancy should be further investigated. Some of the discrepancies can be explained by considering that $\sigma_y \sim a$ and charge collimation was included in the simulations. The difference in the transverse phase space creates a difference in charge transport and charge losses, and therefore the strength of both the longitudinal and transverse wakes.

\section{CLARA FEL Dechirper} \label{sec:CLARADechirper}

\begin{figure}[h]
	\centering
	\includegraphics[width=0.48\linewidth]{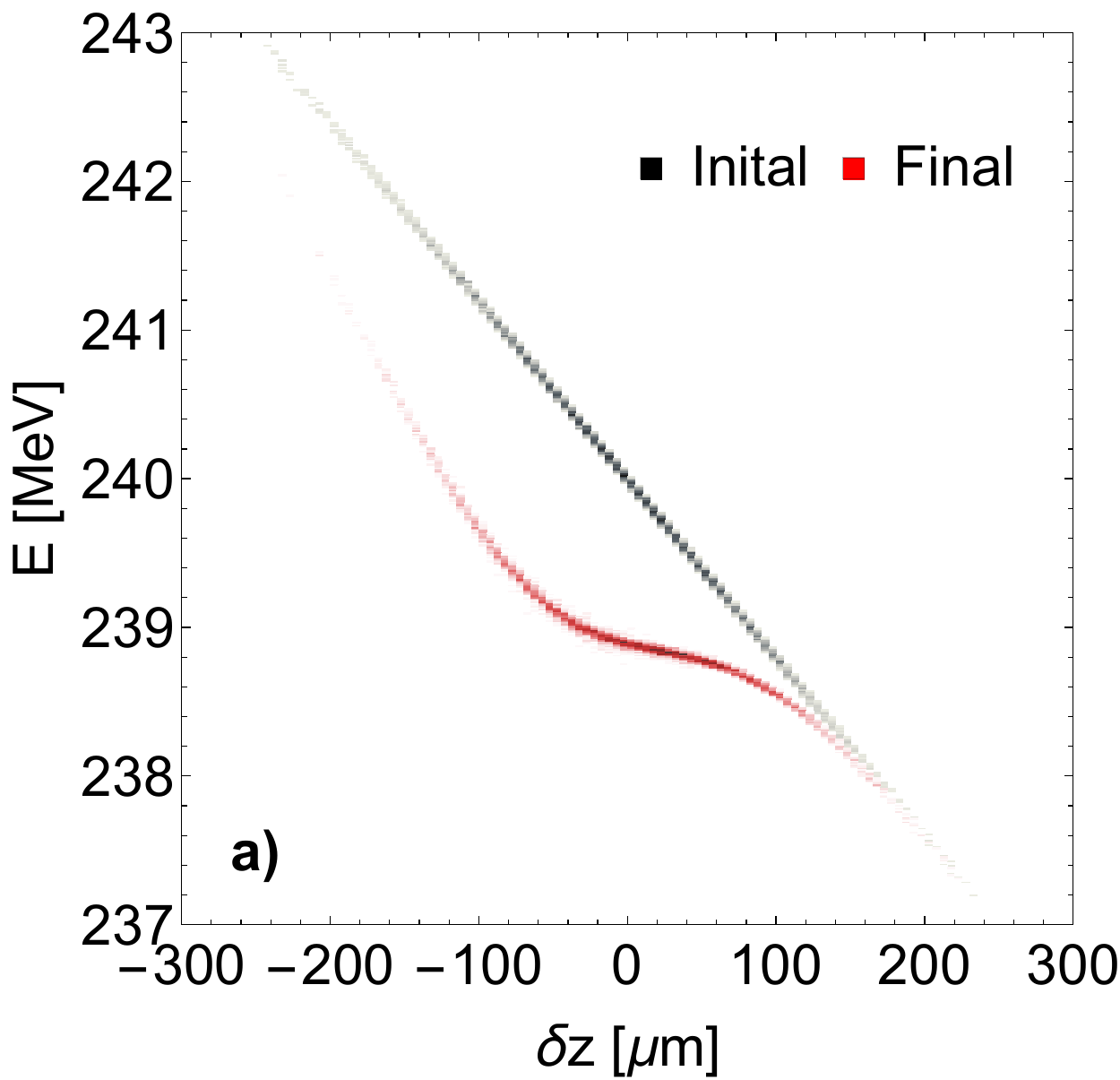}
	\includegraphics[width=0.48\linewidth]{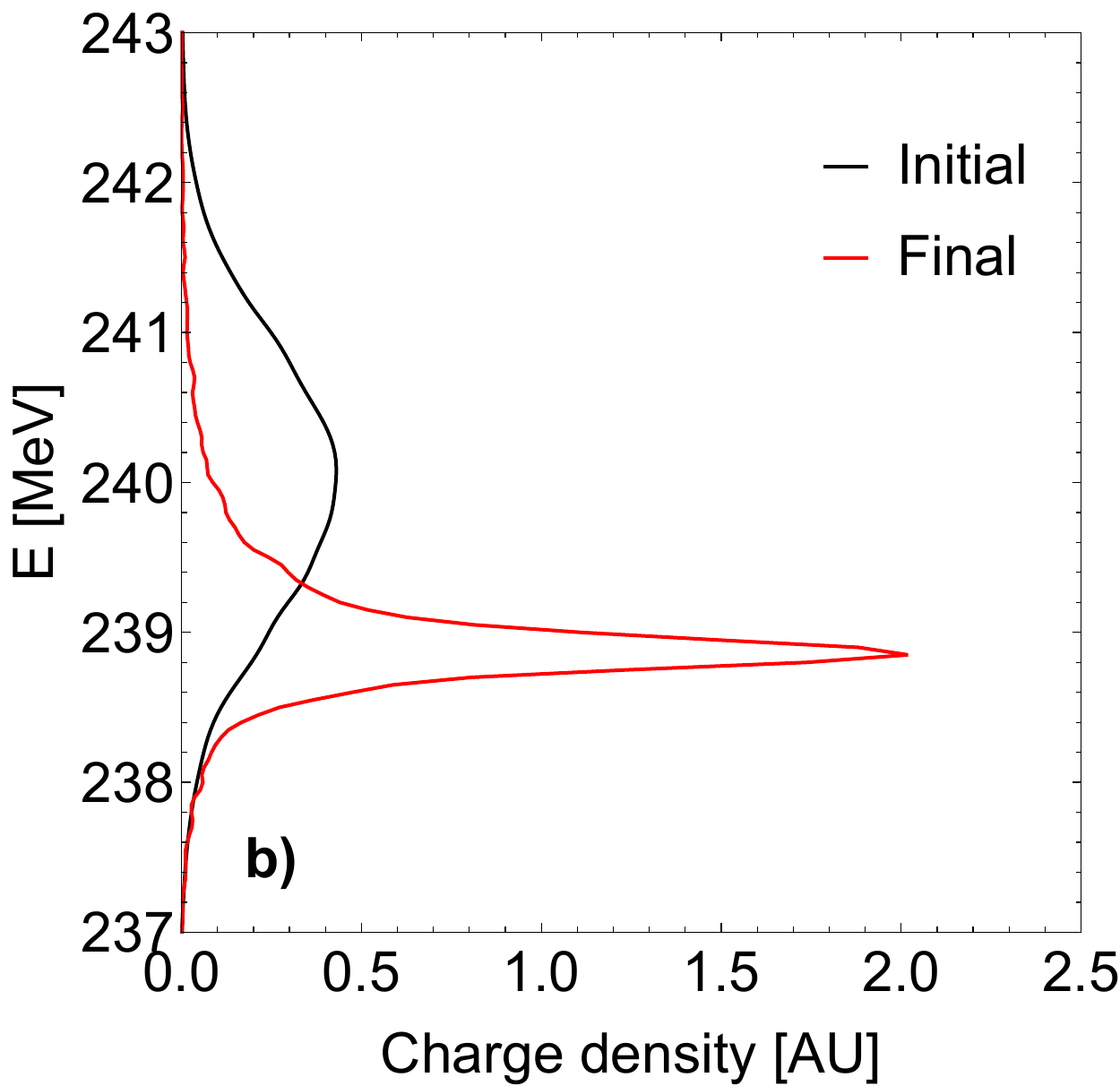}
	\caption{Impact-T simulation for CLARA dechirper, 250 MeV beam, showing flattening of bunch core and energy spread reduction.}
	\label{fig:CLARADechirper}
\end{figure}

Initial feasibility studies of a compact DLW based dechirper for CLARA FEL (beam energy of 250~MeV) have been conducted. The dechirper will comprise of two planar sections, one horizontal and one vertical (H+V), and is similar to the design implemented at LCLS, where corrugated metallic structures were used \cite{guetg2016commissioning}. Such a configuration was demonstrated to preserve transverse emittance of the beam \cite{zhang2015electron}. Results of simulations in Impact-T for a quartz lined structures of combined $10+10$~cm in length, with $\delta = 100$~$\mu$m and $a = 600$~$\mu$m are shown in Figure \ref{fig:CLARADechirper}. The structure reduces the RMS energy spread of the bunch core from 0.24~\% to 0.08~\% and preserves the initial emittance, maintaining $\epsilon_{x,y} < 1$~$\mu$m. Genesis simulations \cite{reiche1999genesis} of the FEL showed pulse power increase of $\sim 30$\%. Further information can be found in \cite{pacey2017phase}. 

\section{Summary}

The dielectric wakefield acceleration programme has been initiated at Daresbury Laboratory with CLARA experimental facility as a testbed. Computer simulations of the planar DLW structure with variable gap and thin 25~$\mu$m dielectric layers confirmed that a tunable narrow bandwidth wakefield generation within 0.5-1.0THz frequency range is achievable. Simulations were performed with two codes, Impact-T and VSim, and a good agreement between the two was demonstrated with regard to longitudinal effects. The difference in simulated transverse effects was more appreciable and this is a subject of further investigation. 
Simulations confirmed that the DLW structure with thin dielectric layers can act as a dechirper for electron bunches with higher electron energy at the head of the bunch i.e. opposite to that used in compression schemes with magnetic chicanes. Feasibility of very compact DLW dechirper for CLARA FEL was also demonstrated.  

\section*{Acknowledgement}
The authors are grateful for the financial support from STFC grant number ST/N504129/1.

\bibliography{references}

\end{document}